\documentclass[a4paper]{article}

\usepackage{INTERSPEECH2021}
\usepackage{amsmath}
\usepackage{graphicx}
\usepackage[acronym]{glossaries}
\usepackage[hyphens]{url}
\usepackage[hidelinks]{hyperref}
\usepackage{tikz}
\usepackage{siunitx}
\usepackage{booktabs}
\usepackage{multirow}
\usepackage{pgfplots}
\usepackage{pgfplotstable}
\usepackage{caption}
\usepackage{subcaption}
\usepackage{pifont}
\usepackage{environ}
\usepackage{balance}

\robustify\bfseries
\usetikzlibrary{calc}
\usetikzlibrary{positioning}
\usetikzlibrary{chains}
\usetikzlibrary{fit}
\usetikzlibrary{arrows}
\usetikzlibrary{decorations.pathreplacing}
\usetikzlibrary{arrows.meta}
\usetikzlibrary{backgrounds}
\usetikzlibrary{tikzmark}
\usetikzlibrary{shapes.multipart}
\usetikzlibrary{matrix}



\tikzset{
	>=stealth,
	element/.style={
		draw=black!100,
		thick,
		align=center,
		inner sep=0pt,
	},
	narrow block/.style={
		element,
		rectangle,
		text width=4.0em,
		minimum width=4.0em,
		minimum height=2.5em,
	},
	block/.style={
		element,
		rectangle,
		text width=7.0em,
		minimum width=7.0em,
		minimum height=2.5em,
	},
	branch/.style={
		element,
		circle,
		fill=black,
		minimum size=0.2em,
	},
	apply/.style={
		element,
		circle,
		minimum size=1em,
		label=center:{$\times$}
	},
	arrow/.style={->, shorten >=0.1em},
	reverse arrow/.style={<-, shorten <=0.1em}
}

\pgfplotsset{compat=1.8}
\usepgfplotslibrary{groupplots}
\pgfplotsset{every tick label/.append style={font=\footnotesize}}

\newcommand{\no}{\ding{56}}
\newcommand{\yes}{\ding{52}}

\newlength{\axisWidth}
\newlength{\axisHeight}

\newacronym{AM}{AM}{acoustic model}
\newacronym{ASR}{ASR}{automatic speech recognition}
\newacronym{DFT}{DFT}{discrete Fourier transform}
\newacronym{KLT}{KLT}{Karhunen-Loève Transform}
\newacronym{LFR}{LFR}{low frame rate}
\newacronym{nWER}{nWER}{normalized word error rate}
\newacronym{WER}{WER}{word error rate}
\newacronym{MVDR}{MVDR}{Minimum Variance Distortionless Response}
\newacronym{RNN-T}{RNN-T}{recurrent neural network transducer}
\newacronym{STFT}{STFT}{short-time Fourier transformation}
\newacronym{VAD}{VAD}{voice activity detection}
\newacronym{VoIP}{VoIP}{voice over IP}
\newacronym{tf-bin}{tf-bin}{time-frequency bin}
\newacronym{SNR}{SNR}{signal-to-noise ratio}
\newacronym{IPD}{IPD}{inter-aural phase difference}
\newacronym{LSTM}{LSTM}{long short-term memory}

\title{Multi-channel Opus compression for far-field automatic speech recognition with a fixed bitrate budget}
\name{Lukas Drude, Jahn Heymann, Andreas Schwarz, Jean-Marc Valin}
\address{Amazon.com}
\email{\{drude, jahheyma, asw, jmvalin\}@amazon.com}

\newcommand{\SIbitrate}[1]{\SI{#1}{kbit/(s\cdot channel)}}
\newcommand{\sibitrate}{\si{kbit/(s\cdot channel)}}

\newwrite\absfile
\NewEnviron{Abstract}{
	\immediate\openout\absfile=\jobname.abs
	\begingroup
	\def\tmp{}\abstractdefs
	\expandafter\stripdollars\BODY$\relax
	\immediate\write\absfile{\tmp}%
	\endgroup
	\abstract\BODY\endabstract
}
\long\def\stripdollars#1${\edef\tmp{\tmp#1}\stripdollarsA}
\def\stripdollarsA{\futurelet\next\stripdollarsB}
\def\stripdollarsB{\ifx\next\relax\else \expandafter\stripdollars\fi}

\def\abstractdefs{
	\def~{ }
	\def\gls##1{##1}
	\def\SI##1##2{##1 ##2}
	\def\second{s}
	\def\par{^^J}
	\def\TeX{TeX}
}

\begin{document}
\ninept
\maketitle
\begin{Abstract}
	\Gls{ASR} in the cloud allows the use of larger models and more powerful multi-channel signal processing front-ends compared to on-device processing. However, it also adds an inherent latency due to the transmission of the audio signal, especially when transmitting multiple channels of a microphone array.
	One way to reduce the network bandwidth requirements is client-side compression with a lossy codec such as Opus.
	However, this compression can have a detrimental effect especially on multi-channel \gls{ASR} front-ends, due to the distortion and loss of spatial information introduced by the codec.
	In this publication, we propose an improved approach for the compression of microphone array signals based on Opus, using a modified joint channel coding approach and additionally introducing a multi-channel spatial decorrelating transform to reduce redundancy in the transmission.
	We illustrate the effect of the proposed approach on the spatial information retained in multi-channel signals after compression, and evaluate the performance on far-field \gls{ASR} with a multi-channel beamforming front-end.
	We demonstrate that our approach can lead to a \SI{37.5}{\percent} bitrate reduction or a \SI{5.1}{\percent} relative \gls{WER} reduction for a fixed bitrate budget in a seven channel setup.
\end{Abstract}


\section{Introduction}
\label{sec:intro}
Far-field \gls{ASR} is machine recognition of speech spoken at a distance from a microphone or microphone array~\cite{haeb2020far}.
This research field has gained popularity due to its widespread application in digital home assistants~\cite{apple2018siri, li2017acoustic, haeb2019speech} and research challenges, e.g.~\cite{barker2017third}.
A common and highly effective approach to improve far-field \gls{ASR} is beamforming \cite{veen1988beamforming}, which requires multiple channels and exploits inter-channel phase and level differences to spatially focus on the desired speaker.

To further improve recognition performance, applications often exploit the computational capabilities of the cloud to run large \gls{ASR} models.
In order to leverage multiple channels in these scenarios, there are two options and both come with different trade-offs:
\begin{enumerate}
	\item The recording device itself is responsible for the signal processing and combines the channels into an enhanced signal which is then transmitted to the cloud for recognition.
	\item The device just records the signal and sends all or selected channels to the cloud for further processing.
\end{enumerate}
The first option minimizes the transmitted amount of data but limits the possible signal processing to the devices compute budget and creates a boundary between the signal processing and acoustic model.
In contrast, the second option does not have this boundary, can combine both stages~\cite{kumatani2019multi, sainath2017multichannel}, and is more flexible in terms of compute budget and methods.
However, transmitting multiple signals requires more bandwidth which is often very limited for a large number of users.
In order to save bandwidth, the device can compress the signals before transmission.
The challenge in the compression of multi-channel signals is that the data rate needs to be sufficiently reduced to prevent impact on the latency perceived by the user, while ensuring that spatial information is preserved in order to allow cloud-side signal enhancement to be effective.

We here focus on Opus compression~\cite{valin2012rfc} due to its widespread use in \gls{VoIP} and \gls{ASR} applications~\cite{valin2016webrtc}.
While Opus has been found to be effective for single- and multi-channel \gls{ASR} applications \cite{khare2020multichannel}, it is in some sense mismatched to the task at hand.
First, like all lossy speech and audio compression methods, Opus has been optimized for subjective quality, not for \gls{ASR} or subsequent spatial filtering.
Second, Opus does not have dedicated support for efficient compression of highly correlated multi-channel signals.

Quite a few studies assess Opus compression in terms of human perception, e.g., in~\cite{orosz2013performance} the authors perform perceptual listening tests when Opus is applied in a \gls{VoIP} setting but do not address the far-field \gls{ASR} setup.
A perceptual quality assessment in terms of speech quality and localization accuracy was performed in~\cite{narbutt2017streaming}, where ambisonic representations were compressed with Opus where the authors mainly experimented with different bitrates but did not analyze coding features more targeted to spatial fidelity.

With focus on \gls{ASR} applications, \cite{narayanan2018toward} analyzes the impact of different compression algorithms on a single-channel \gls{AM} and concludes that using a compression algorithms as a data augmentation step in the training phase of an \gls{AM} can lead to a more robust model.
Closely related to the document at hand, Khare et al. analyzed the impact of Opus compression on multi-channel \glspl{AM}~\cite{khare2020multichannel} with the main findings that multi-channel data is preferable over single-channel transmission with a higher bitrate.
They further report gains when retraining the \gls{AM}, which we do not repeat here.
However, the authors leave it open how compression impacts a beamforming front-end and provide little insight into the spatial properties of the compressed signals.
Their investigated \gls{ASR} system combines the multi-channel processing and senone classification into one model capable of incorporating compression artifacts during training as their experiments in Section~5 show.
In contrast, analytic beamforming, as analyzed here, relies on unaltered phase information and, therefore, requires different treatment.

This contribution analyzes the impact of Opus compression on a combined beamforming and \gls{ASR} system, provides guidelines on optimizing compression parameters for far-field \gls{ASR}, and proposes a microphone-independent multi-channel coding scheme.

\begin{figure*}
	\centering
	\input{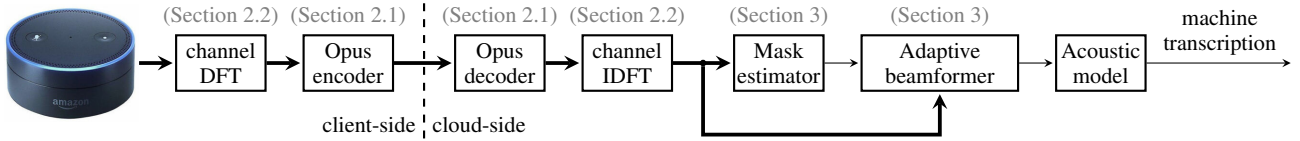}
	\vspace{-6mm}
	\caption{Overview of a beamforming-based far-field \gls{ASR} system.
		Compression is applied to the multi-channel signal to enable all other signal processing in the cloud.
		The visualization contains the proposed \acrshort{DFT} across channels as a transform coding scheme.}
	\label{figure:overview}
\end{figure*}

\section{System description}
\label{sec:format}
The overall far-field \gls{ASR} pipeline including a neural mask-based beamforming front-end is illustrated in Figure~\ref{figure:overview}:
A microphone array (here consisting of 7 channels) records far-field audio impaired by reverberation, background noise, and potentially non-device-directed speech.
All channels are then compressed as detailed in Section~\ref{sec:opus}.
Section~\ref{sec:dft} details the proposed \gls{DFT}-based coding scheme.
A neural mask estimator or a neural \gls{VAD} is used to control an analytic beamformer (see Section~\ref{sec:setup}).
Finally, transcriptions are obtained with a \gls{RNN-T} \gls{AM} (see Section~\ref{sec:setup})~\cite{graves2012sequence}.

\subsection{Opus compression and proposed optimization}
\label{sec:opus}
The Opus codec aims at real-time speech or music applications~\cite{valin2012rfc, valin2013high}.
It originates from Skype’s SILK codec~\cite{vos2013voice}, which relies on linear predictive coding~\cite{makhoul1975lpc} and Xiph.Org's CELT codec~\cite{valin2009full}, which uses a modified discrete cosine transform.
The Opus codec allows to encode with SILK, with CELT, or with a hybrid of both within which lower frequencies dominated by speech are using SILK while higher frequencies fall back to CELT with specifics depending on selected parameters such as bitrate, the number of channels, or the audio content~\cite{valin2013high}.
The encoder can then dynamically decide between both modes.
Depending on externally provided media type flags one can bias these decisions towards either SILK or CELT.

While Opus has been found to be effective for single- and multi-channel \gls{ASR} applications \cite{khare2020multichannel}, it has been optimized for subjective quality, not for \gls{ASR} accuracy.
Moreover, its channel coupling has been optimized for human spatial perception, where the \gls{IPD} \cite{moore2012introduction} is only noticeable at lower frequencies.
This is why the CELT coding mode may make use of intensity stereo past a certain bitrate-dependent threshold frequency.
Frequency bands coded with intensity stereo share the same phase and spectral details and only differ between channels by the energy (inter-aural intensity difference).
Thus, intensity stereo can severely degrade the performance of beamforming algorithms, which rely on accurate phases.
Similarly, when not enough bits are available, Opus relies on spectral folding to \emph{fill} spectral bands with perceptually acceptable content.
Again, while it is a good perceptual trade-off (allowing for more bits at lower frequency), the feature impairs beamforming algorithms.

Opus does not have dedicated multi-channel support.
However, it supports joint compression of two channels~\cite{valin2012rfc}:
In the SILK coding algorithm -- optimized specifically for voice -- coupled channels are split into a mid (average) and a side (difference) signal to decorrelate the channels.
While the decomposition preserves the phase, the encoder will sometimes choose to discard the spatial information (side signal) in favor of reducing perceptible coding noise in the mid signal.
This destroys useful phase information for a beamformer.

For these reasons, we propose to disable CELT intensity stereo and reduce the use of folding to the minimum permitted by the Opus bitstream.
For SILK, the bit allocation split between mid and side is set to optimize phase encoding rather than perceptual quality.
In the following this is denoted as \emph{waveform matching}\footnote{\url{https://gitlab.xiph.org/xiph/opus/-/commit/cece1fd6}}.
Since this option only changes the encoder, the result can still be decoded with any compliant decoder.

\subsection{Proposed transform coding scheme}
\label{sec:dft}
Motivated by the high correlation between the channels in a closely spaced microphone array, one may look for an orthogonal decomposition to avoid coding redundant information. For spherical or hemispherical microphone arrays, orthogonal decompositions are known~\cite{rafaely2004analysis, li2005hemispherical}.
However, given that all other signal components in this work are independent of the microphone array geometry, such additional restrictions on array geometry are undesired.
Another option is spatial whitening~\cite{venkatesan2004iterative} which can be achieved by a \gls{KLT} \cite{yang2003high}.
However, this transform is signal-dependent and requires the estimation of a spatial correlation matrix and potentially updating it online due to the nonstationary nature of the spatial signal characteristics.

To avoid the aforementioned drawbacks while yielding some decorrelation, we propose to apply a $D$-point \gls{DFT} across channels (note that this is not a DFT along time):
\begin{align}
	X_k(t) = \sum_{d=0}^{D-1} x_d(t) \cdot \mathrm{e}^{-\mathrm{j}2\pi d k / D}, \label{eq:dft}\vspace{-2ex}
\end{align}
where $t$ is the time index, $d$ is the channel index, and $k$ is the \gls{DFT} bin index.
In contrast to spherical decomposition and spatial whitening this approach does not make strong assumptions on the array geometry and does not depend on the signal statistics themselves, yet still achieves a significant amount of decorrelation (compare Table~\ref{table:power} for a power distribution over channels).
Since the \gls{DFT} of a real-valued signal is even symmetric, i.e., $X_k(t) = X^{\ast}_{-k\,\mathrm{mod}\,D}(t)$, we effectively obtain a real-valued DC component $X_0(t)$ and, when $D$ is odd, $(D - 1) / 2$ complex-valued non-redundant channels.
This can be seen as a generalization of mid/side stereo to an arbitrary number of channels.
Further, it can be interpreted as an approximation of the \gls{KLT} coding scheme \cite{yang2003high}.
Finally, after Opus decoding, the original channels can be reconstructed using the corresponding inverse \gls{DFT} over channels corresponding to Equation~\ref{eq:dft}.

In the experimental section we further demonstrate that the joint compression of the real and imaginary parts of one \gls{DFT} bin, i.e., $\mathrm{Re}\{X_d\}$ and $\mathrm{Im}\{X_d\}$, is beneficial.
\begin{table}[h]
	\caption{Signal power distribution across physical and \acrshort{DFT} channels. All values are averaged over 1000 examples and listed in \si{\percent}.}
	\label{table:power}
	\centering
	\begin{tabular}{
			l
			S[table-format=2.1, round-mode=places, round-precision=1]@{\hskip0.4em}
			S[table-format=2.1, round-mode=places, round-precision=1]@{\hskip0.5em}
			S[table-format=2.1, round-mode=places, round-precision=1]@{\hskip0.5em}
			S[table-format=2.1, round-mode=places, round-precision=1]@{\hskip0.5em}
			S[table-format=2.1, round-mode=places, round-precision=1]@{\hskip0.5em}
			S[table-format=2.1, round-mode=places, round-precision=1]@{\hskip0.5em}
			S[table-format=2.1, round-mode=places, round-precision=1]@{\hskip0.5em}
		}
		\toprule
		Index & {0} & {1} & {2} & {3} & {4} & {5} & {6} \\
		\midrule
		Physical channels & 13.51378858089447 & 14.092603325843811 & 13.6461541056633 & 15.625107288360596 & 16.415342688560486 & 14.611464738845825 & 12.095534801483154 \\
		\acrshort{DFT} channels & 87.09610136430797 & 3.948420670020996 & 1.4299952073865878 & 1.07353344043843 & 1.07353344043843 & 1.4299952073865878 & 3.948420670020996 \\
		\bottomrule
	\end{tabular}
\end{table}

\begin{figure*}
	\setlength{\axisWidth}{5.75cm}
	\setlength{\axisHeight}{4cm}
	\centering
	\input{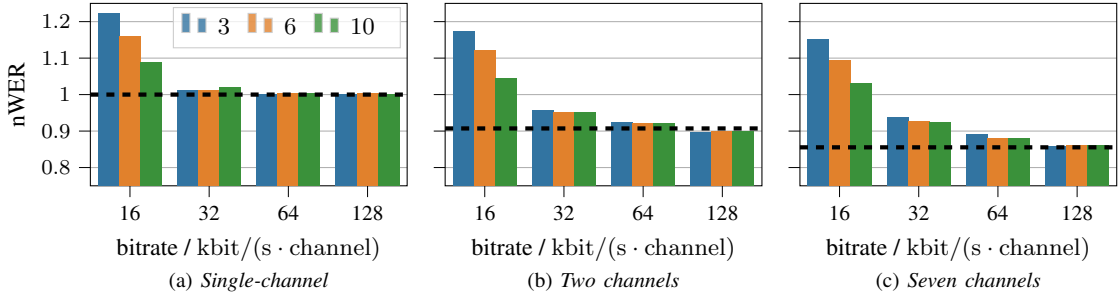}
	\vspace{-4mm}
	\caption{Comparison of different compression complexities (in colors), different bitrates (groups in each plot) for single-channel (first plot), two channels (second plot), and seven channels (third plot) in terms of \acrshort{nWER}.
		All results are shown with automatic compression mode selection and channel-independent compression.
		The corresponding uncompressed baseline is shown as a black dashed line.}
	\label{figure:bars}
\end{figure*}

\pagebreak
\begin{table}[b]
	\caption{Comparison of channel-independent vs. pairwise compression.
		All compression results are with CELT compression.}
	\label{table:pairwise}
	\centering
	\begin{tabular}{S[table-format=3]@{\hskip0.75em}S[table-format=1]@{\hskip0.75em}S[table-format=1.3]@{\hskip0.75em}S[table-format=1.3]S[table-format=-1.1]}
		\toprule
		{\multirow{2.5}{*}{\shortstack{Bitrate /\\\sibitrate}\!\!\!\!}} &
		{\multirow{2.5}{*}{\shortstack{Channels\\$N$}}} & \multicolumn{2}{c}{\acrshort{nWER}} & {\multirow{2.5}{*}{\shortstack{Delta /\\\si{\percent}}}} \\
		\cmidrule{3-4}
		&& {indep.} & {pairwise} & \\
		\midrule
		{uncompressed} & 2 & \multicolumn{2}{c}{\num{0.907}} & \\
		\midrule
		16 & 2 & 1.018 & 1.010 & -0.8 \\
32 & 2 & 0.949 & 0.933 & -1.7 \\
64 & 2 & 0.923 & 0.910 & -1.4 \\
128 & 2 & 0.901 & 0.899 & -0.1 \\
		\bottomrule
	\end{tabular}
\end{table}
\section{Experimental setup}
\label{sec:setup}
All experiments are performed on \SI{16}{kHz} de-identified audio with up to seven channels.
Typical microphone arrays for \gls{ASR} consist of sensors spaced only few centimeters apart.
The recorded signals are therefore highly correlated and differ primarily in phase.
Beamforming~\cite{veen1988beamforming} uses this phase difference to spatially filter the signal in order to suppress undesired components while enhancing the desired signal resulting in fewer recognition errors~\cite{Compernolle1990}.
Here, we consider neural mask-based beamforming where the information about the dominant source is estimated on a \gls{tf-bin} level using a combination of a neural \gls{VAD} and a neural mask estimator~\cite{heymann2016neural, erdogan2016improved}.
The mask estimator is trained on separate synthetic data and is supported by a neural \gls{VAD}~\cite{zhang2012deep} trained using production data and a forced alignment as reference.
The beamforming vector is obtained using a \gls{MVDR} formulation as in~\cite{Souden2013} with a fixed reference channel.

We use a \gls{RNN-T} single-channel acoustic model~\cite{graves2012sequence}.
It consists of an eight \gls{LSTM} layer encoder and a two \gls{LSTM} layer decoder.
Each \gls{LSTM} layer has \(1024\) units.
The encoder and decoder outputs are projected to a \(512\) dimensional vector which is combined by the joint network with a single layer with \(1024\) units to provide a distribution over \SI{4}{k} wordpieces.
It is trained with SpecAugment \cite{Park2020} on $\approx$\SI{60000}{h} of de-identified production data, which is the single-channel output of an on-device beamformer.
The \gls{RNN-T} training data has been Opus-compressed with default settings and bitrates between 32 and \SIbitrate{64}.

For evaluation we use a de-identified dataset consisting of \SI{126}{h} of realistic production-like speech recorded by Amazon Echo-style microphone arrays.
The dataset is available as uncompressed audio, which allows to conduct decoding experiments with different lossy compression setups against an uncompressed baseline.
All experimental results are presented in terms of \gls{nWER} such that the single-channel baseline (single digit in percent) without any encoding results in a \gls{nWER} of \num{1.}.

\section{Experimental Results}
Figure~\ref{figure:bars} visualizes the impact of Opus compression on a single-channel \gls{AM} (\ref{subplot:single}), a two-channel beamforming front-end with a single-channel \gls{AM} (\ref{subplot:two}) and a seven-channel beamforming front-end with a single-channel \gls{AM} (\ref{subplot:seven}).
The compression mode is selected automatically (auto).
The main observation is, that the \gls{nWER} already saturates for about \SIbitrate{32} for the single-channel \gls{AM}, while all beamforming variants still heavily degrade with bitrates below \SIbitrate{128}.
Further, we can deduce that especially for low bitrates a high compression complexity (here shown in colors) improves \glspl{WER} significantly.

Table~\ref{table:pairwise} compares pairwise compression using the proposed waveform matching with the channel-independent compression for different bitrates.
It turns out that for all bitrates pairwise compression with waveform matching performs better with increased gains of up to \SI{1.7}{\percent} relative for realistic bitrates.

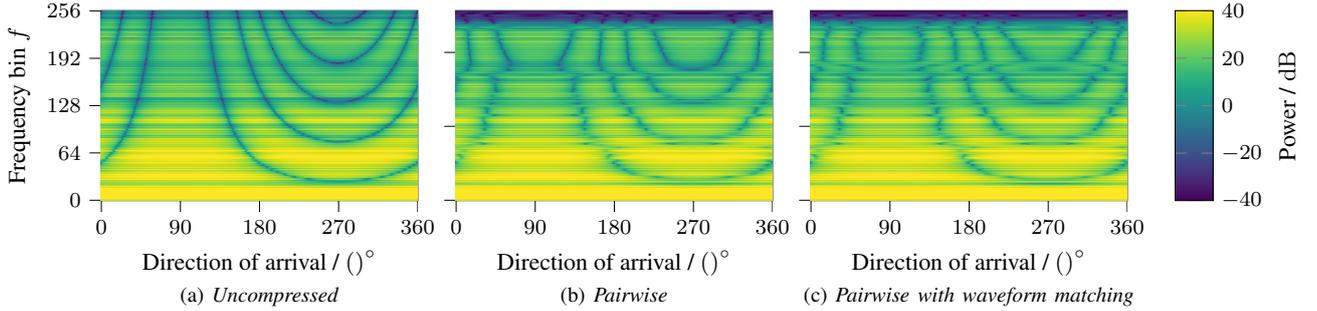
\begin{figure*}
	\setlength{\axisWidth}{5.75cm}
	\setlength{\axisHeight}{4.1cm}
	\centering
\begin{tikzpicture}

\begin{groupplot}[group style={group name=my plots, horizontal sep=0.5cm, group size=3 by 1}]
\nextgroupplot[
height=\axisHeight,
tick align=outside,
tick pos=left,
width=\axisWidth,
x grid style={white!69.0196078431373!black},
xlabel={Direction of arrival / \(\displaystyle ()^{\circ}\)},
xtick = {0, 90, 180, 270, 360},
xmin=-0.5, xmax=360,
xtick style={color=black},
y grid style={white!69.0196078431373!black},
ylabel={Frequency bin \(\displaystyle f\)},
ymin=-0.5, ymax=256.5,
ytick={0, 64, 128, 192, 256},
ytick style={color=black}
]
\addplot graphics [includegraphics cmd=\pgfimage,xmin=-0.5, xmax=359.5, ymin=-0.5, ymax=256.5] {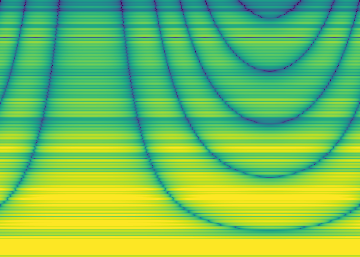};

\nextgroupplot[
height=\axisHeight,
scaled y ticks=manual:{}{\pgfmathparse{#1}},
tick align=outside,
tick pos=left,
width=\axisWidth,
x grid style={white!69.0196078431373!black},
xlabel={Direction of arrival / \(\displaystyle ()^{\circ}\)},
xtick = {0, 90, 180, 270, 360},
xmin=-0.5, xmax=360,
xtick style={color=black},
y grid style={white!69.0196078431373!black},
ymin=-0.5, ymax=256.5,
ytick style={color=black},
yticklabels={}
]
\addplot graphics [includegraphics cmd=\pgfimage,xmin=-0.5, xmax=359.5, ymin=-0.5, ymax=256.5] {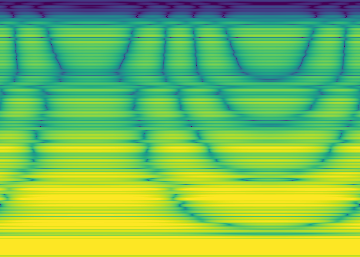};

\nextgroupplot[
colorbar,
colorbar style={ylabel={Power / $\mathrm{dB}$}},
colormap/viridis,
height=\axisHeight,
point meta max=40,
point meta min=-40,
scaled y ticks=manual:{}{\pgfmathparse{#1}},
tick align=outside,
tick pos=left,
width=\axisWidth,
x grid style={white!69.0196078431373!black},
xlabel={Direction of arrival / \(\displaystyle ()^{\circ}\)},
xtick = {0, 90, 180, 270, 360},
xmin=-0.5, xmax=360,
xtick style={color=black},
y grid style={white!69.0196078431373!black},
ymin=-0.5, ymax=256.5,
ytick style={color=black},
yticklabels={}
]
\addplot graphics [includegraphics cmd=\pgfimage,xmin=-0.5, xmax=359.5, ymin=-0.5, ymax=256.5] {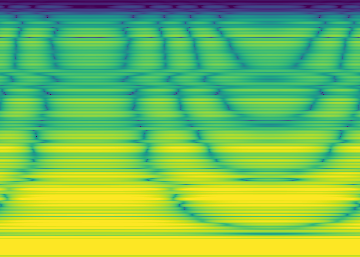};
\end{groupplot}
\node[text width=6cm,align=center,anchor=north] at ([yshift=-8mm]my plots c1r1.south) {\subcaption{Uncompressed\label{subplot:uncompressed}}};
\node[text width=6cm,align=center,anchor=north] at ([yshift=-8mm]my plots c2r1.south) {\subcaption{Pairwise\label{subplot:pw}}};
\node[text width=6cm,align=center,anchor=north] at ([yshift=-8mm]my plots c3r1.south) {\subcaption{Pairwise with waveform matching\label{subplot:pw_wm}}};

\end{tikzpicture}
	\vspace{-7mm}
	\caption{Visualization of the impact of pairwise compression on the spatial characteristics of an artificially delayed recording in terms of a direction-of-arrival spectrum.
		A fixed steering vector for a delay-and-sum beamformer is calculated for each look direction. The response power for each direction is then a visualization of the spatial properties of the  signal.}
	\label{figure:doa}
\end{figure*}

\begin{table}[b]
	\caption{Ablation study for a given bitrate of \SIbitrate{32} in a two-channel beamforming system in terms of \acrshort{nWER}.}
	\label{table:ablation}
	\centering
	\begin{tabular}{lS[table-format=1]cS[table-format=1.3]S[table-format=1.3]}
		\toprule
		\multirow{2.5}{*}{\shortstack{Compression\\mode}} &
		{\multirow{2.5}{*}{\shortstack{Channels\\$N$}}} &
		\multirow{2.5}{*}{\shortstack{Waveform\\matching}\!\!} &
		\multicolumn{2}{c}{\acrshort{nWER}} \\
		\cmidrule{4-5}
		& & & {indep.} & {pairwise} \\
		\midrule
		uncompressed\hspace{-2em} & 2 & & \multicolumn{2}{c}{\num{0.907}} \\  %
		\midrule
		SILK & 2 & \no & 0.960 & 0.933 \\
SILK & 2 & \yes & 0.962 & 0.933 \\
auto & 2 & \no & 0.952 & 0.959 \\
auto & 2 & \yes & 0.957 & 0.933 \\
CELT & 2 & \no & 0.949 & 0.959 \\
CELT & 2 & \yes & 0.949 & 0.933 \\
		\bottomrule
	\end{tabular}
\end{table}
\begin{table}[b]
	\caption{Comparison of channel-independent CELT compression and transform coding using a channel-wise \acrshort{DFT} before CELT compression and a channel-wise inverse \acrshort{DFT} after decompression.
		The delta values in this table are w.r.t. the channel-independent column.}
	\label{table:dft}
	\centering
	\begin{tabular}{S[table-format=3]S[table-format=1]S[table-format=1.3]S[table-format=1.3]S[table-format=-1.1]}
		\toprule
		{\multirow{2.5}{*}{\shortstack{Bitrate /\\\sibitrate}}\!\!\!\!} &
		{\multirow{2.5}{*}{\shortstack{Channels\\$N$}}} &
		\multicolumn{2}{c}{\acrshort{nWER}} &
		{\multirow{2.5}{*}{\shortstack{Delta /\\\si{\percent}}}} \\
		\cmidrule{3-4}
		& & {indep.} & {\acrshort{DFT}} & \\
		\midrule
		16 & 7 & 1.015 & 0.947 & -6.7 \\
32 & 7 & 0.925 & 0.888 & -4.0 \\
64 & 7 & 0.882 & 0.870 & -1.4 \\
128 & 7 & 0.861 & 0.853 & -0.9 \\
		\bottomrule
	\end{tabular}
\end{table}

Table~\ref{table:ablation} puts the compression mode, waveform matching and pairwise compression into relation.
For channel-independent compression, CELT compression performs best independently of the proposed waveform matching.
However, when channels are compressed in pairs the proposed waveform matching tends to reduce \glspl{WER}.

To understand the impact of Opus compression on spatial properties we took a single \SI{7.4}{s} speech recording and delayed a copy by $\Delta=5$ samples.
Then, we calculated a direction-of-arrival spectrum using a fixed delay-and-sum beamformer for each look direction.
Figure~\ref{subplot:uncompressed} shows the direction-of-arrival spectrum for the uncompressed two-channel signal.
For all frequencies, the signals positively interfere at around \SI{90}{\degree}.
U-shaped regions where signals cancel out are clearly visible.
For pairwise compression with a bitrate of \SI{32}{kbit/(s \cdot channel)}, Figure~\ref{subplot:pw} shows that the curves of signal cancellation are much more distorted.
Most prominently, the phase information is entirely lost for frequency bin indices $f>180$.
Figure~\ref{subplot:pw_wm} shows results with the proposed waveform matching, demonstrating that the loss of phase information is avoided.
Given the limited bitrate this likely resulted in reduced perceptual quality.
However, with a beamforming application in mind, correct spatial properties are preferred over perceptual fidelity.

Finally, we evaluate the aforementioned improvements in combination with our proposed \gls{DFT}-based coding scheme on a seven-channel microphone array signal.
Table~\ref{table:dft} shows the \glspl{nWER} and the corresponding relative \gls{WER} changes (delta).
The ``indep.'' column shows \glspl{nWER} for independent coding of channels.
The ``DFT'' column shows \glspl{nWER} for signals coded with the proposed decorrelation using a \gls{DFT} across channels.
After Opus decoding the signal is reconstructed using an inverse \gls{DFT} over channels.
Transform coding improves \gls{WER} for all bitrates with relative gains up to \SI{6.7}{\percent}%
, with the largest gain observed for the lowest bitrates, highlighting its importance for cohorts with limited internet connectivity.
Alternatively, instead of optimizing for \gls{WER} we can fix the \gls{WER} at the value obtained for \SIbitrate{32} channel-independent compression.
Using the proposed channel decorrelation we can reduce the bitrate down to \SIbitrate{20} without compromising on \gls{WER}.
This corresponds to a bitrate reduction of \SI{37.5}{\percent}.
These findings encourage further research towards multi-channel extensions for lossy compression.

After applying a \gls{DFT} across seven channels one obtains a real-valued channel (sometimes called the DC component) and three distinct complex-valued channels.
Under the assumption that transformed channels and their respective quantization noise are mutually uncorrelated, transform coding principles suggest that bitrate should be allocated relative to signal power of each channel (compare Table~\ref{table:power} for a power distribution across channels).
However, Table~\ref{table:transform}, in which we experimented with increasing the bitrate for the DC channel which has the highest power while keeping the average bitrate constant, shows, that uniform bitrate assignment is better in our case.
Additional experiments with signal power-based bitrate assignments resulted in worse \gls{WER} and are omitted here for brevity reasons.

Given that the real and imaginary part of one \gls{DFT} channel show similar dynamics, one may conclude that pairwise compression may be beneficial to allow the Opus codec to better exploit residual redundancies between those parts.
Indeed, as indicated by the ``\acrshort{DFT}+pairwise'' column in Table~\ref{table:transform}, larger \gls{WER} changes can be achieved with additional pairwise compression.
Here, we also use the proposed waveform matching for the complex pairs.
\begin{table}[h]
	\caption{%
		Relative \gls{WER} change of the proposed transform coding over channel-independent compression including all prior optimizations.
		The ``\acrshort{DFT}'' column corresponds to the \gls{DFT} across channels while the ``\acrshort{DFT}+pairwise'' column corresponds to additional pairwise compression of the real and imaginary part of each AC component, i.e., $\mathrm{Re}\{X_d\}$ and $\mathrm{Im}\{X_d\}$ for $d \in {1,\dots D-1}$.
	}
	\label{table:transform}
	\centering
	\begin{tabular}{%
			c%
			c%
			S[table-format=-1.1, round-mode=places, round-precision=1]@{\hskip0.5em}%
			S[table-format=-1.1, round-mode=places, round-precision=1]%
		}
		\toprule
		{\multirow{2.5}{*}{\shortstack{Bitrate /\\\sibitrate}}\!\!\!\!} &
		\multicolumn{3}{c}{\shortstack{Delta / \si{\percent}}} \\
		\cmidrule{2-4}
		& {indep.} & {\acrshort{DFT}} & {\acrshort{DFT}+pairwise} \\
		\midrule
		All channels 32 & $\pm 0.0$ & -4.01 & \bfseries -5.12 \\
DC channel 38, others 31 & & -3.45 & -4.81 \\
DC channel 44, others 30 & & -3.65 & -4.48 \\
DC channel 50, others 29 & & -3.29 & -4.88 \\
DC channel 56, others 28 & & -3.37 & -4.41 \\
DC channel 62, others 27 & & -3.32 & -3.91 \\
		\bottomrule
	\end{tabular}
\end{table}

\section{Conclusions}
We analyzed why multi-channel far-field \gls{ASR} degrades under Opus compression with limited bitrates and proposed modifications to the codec in order to reduce \gls{WER} while maintaining compatibility with existing decoders.
In a two-channel beamforming application this leads to up to \SI{2}{\percent} relative \gls{WER} reduction without any increase in bitrate. %
We furthermore propose a new spatial transform coding scheme-based on a \gls{DFT} decomposition across channels.
While it is a simple operation it maintains compatibility to existing Opus encoder/decoder infrastructure, and is geometry- and signal-independent.
For a seven-channel beamforming application it allows to either reduce the bitrate by \SI{37.5}{\percent} at no loss in accuracy or alternatively reduce \gls{WER} by \SI{5.1}{\percent} relative with a fixed bitrate budget.
Overall, our proposed multi-channel coding scheme achieves a relative \gls{WER} reduction of \SI{6.4}{\percent} for a seven-channel microphone array compared to a naive channel-independent Opus compression.
\vfill\pagebreak\clearpage
\balance
\bibliographystyle{IEEEtran}
\bibliography{refs}

\begin{thebibliography}{10}
\providecommand{\url}[1]{#1}
\csname url@samestyle\endcsname
\providecommand{\newblock}{\relax}
\providecommand{\bibinfo}[2]{#2}
\providecommand{\BIBentrySTDinterwordspacing}{\spaceskip=0pt\relax}
\providecommand{\BIBentryALTinterwordstretchfactor}{4}
\providecommand{\BIBentryALTinterwordspacing}{\spaceskip=\fontdimen2\font plus
\BIBentryALTinterwordstretchfactor\fontdimen3\font minus
  \fontdimen4\font\relax}
\providecommand{\BIBforeignlanguage}[2]{{%
\expandafter\ifx\csname l@#1\endcsname\relax
\typeout{** WARNING: IEEEtran.bst: No hyphenation pattern has been}%
\typeout{** loaded for the language `#1'. Using the pattern for}%
\typeout{** the default language instead.}%
\else
\language=\csname l@#1\endcsname
\fi
#2}}
\providecommand{\BIBdecl}{\relax}
\BIBdecl

\bibitem{haeb2020far}
R.~Haeb-Umbach, J.~Heymann, L.~Drude, S.~Watanabe, M.~Delcroix, and
  T.~Nakatani, ``Far-field automatic speech recognition,'' \emph{Proceedings of
  the IEEE}, 2020.

\bibitem{apple2018siri}
{Audio Software Engineering and Siri Speech Team}, ``Optimizing {Siri} on
  {HomePod} in far‑field settings,''
  \url{https://machinelearning.apple.com/research/optimizing-siri-on-homepod-in-far-field-settings},
  2018.

\bibitem{li2017acoustic}
B.~Li, T.~N. Sainath, A.~Narayanan, J.~Caroselli, M.~Bacchiani, A.~Misra,
  I.~Shafran, H.~Sak, G.~Pundak, K.~K. Chin \emph{et~al.}, ``Acoustic modeling
  for {Google} {Home},'' in \emph{Interspeech}, 2017.

\bibitem{haeb2019speech}
R.~Haeb-Umbach, S.~Watanabe, T.~Nakatani, M.~Bacchiani, B.~Hoffmeister, M.~L.
  Seltzer, H.~Zen, and M.~Souden, ``Speech processing for digital home
  assistants: Combining signal processing with deep-learning techniques,''
  \emph{IEEE Signal processing magazine}, vol.~36, no.~6, pp. 111--124, 2019.

\bibitem{barker2017third}
J.~Barker, R.~Marxer, E.~Vincent, and S.~Watanabe, ``The third {CHiME} speech
  separation and recognition challenge: Analysis and outcomes,'' \emph{Computer
  Speech \& Language}, vol.~46, pp. 605--626, 2017.

\bibitem{veen1988beamforming}
B.~D. {Van Veen} and K.~M. {Buckley}, ``Beamforming: a versatile approach to
  spatial filtering,'' \emph{IEEE ASSP Magazine}, vol.~5, no.~2, pp. 4--24,
  1988.

\bibitem{kumatani2019multi}
K.~Kumatani, W.~Minhua, S.~Sundaram, N.~Str{\"o}m, and B.~Hoffmeister,
  ``Multi-geometry spatial acoustic modeling for distant speech recognition,''
  in \emph{International Conference on Acoustics, Speech and Signal Processing
  (ICASSP)}.\hskip 1em plus 0.5em minus 0.4em\relax IEEE, 2019.

\bibitem{sainath2017multichannel}
T.~N. Sainath, R.~J. Weiss, K.~W. Wilson, B.~Li, A.~Narayanan, E.~Variani,
  M.~Bacchiani, I.~Shafran, A.~Senior, K.~Chin \emph{et~al.}, ``Multichannel
  signal processing with deep neural networks for automatic speech
  recognition,'' \emph{IEEE/ACM Transactions on Audio, Speech, and Language
  Processing}, vol.~25, no.~5, 2017.

\bibitem{valin2012rfc}
J.-M. Valin, K.~Vos, and T.~B. Terriberry, ``Definition of the {Opus} audio
  codec,'' \url{https://tools.ietf.org/html/rfc6716}, pp. 1--326, September
  2012.

\bibitem{valin2016webrtc}
J.-M. Valin and C.~Bran, ``{WebRTC} audio codec and processing requirements,''
  \url{https://tools.ietf.org/html/rfc7874}, 2016.

\bibitem{khare2020multichannel}
A.~Khare, S.~Sundaram, and M.~Wu, ``Multi-channel acoustic modeling using mixed
  bitrate {Opus} compression,'' \emph{arXiv preprint arXiv:2002.00122}, 2020.

\bibitem{orosz2013performance}
P.~Orosz, T.~Skopk{\'o}, Z.~Nagy, and T.~Lukovics, ``Performance analysis of
  the {Opus} codec in {VoIP} environment using {QoE} evaluation,'' in
  \emph{International Conference on Systems and Networks Communications
  (ICSNC)}, 2013.

\bibitem{narbutt2017streaming}
M.~Narbutt, S.~O'Leary, A.~Allen, J.~Skoglund, and A.~Hines, ``Streaming {VR}
  for immersion: Quality aspects of compressed spatial audio,'' in
  \emph{International Conference on Virtual System \& Multimedia (VSMM)}.\hskip
  1em plus 0.5em minus 0.4em\relax IEEE, 2017.

\bibitem{narayanan2018toward}
A.~Narayanan, A.~Misra, K.~C. Sim, G.~Pundak, A.~Tripathi, M.~Elfeky,
  P.~Haghani, T.~Strohman, and M.~Bacchiani, ``Toward domain-invariant speech
  recognition via large scale training,'' in \emph{Spoken Language Technology
  Workshop (SLT)}.\hskip 1em plus 0.5em minus 0.4em\relax IEEE, 2018.

\bibitem{graves2012sequence}
A.~Graves, ``Sequence transduction with recurrent neural networks,''
  \emph{arXiv preprint arXiv:1211.3711}, 2012.

\bibitem{valin2013high}
J.-M. Valin, G.~Maxwell, T.~B. Terriberry, and K.~Vos, ``High-quality,
  low-delay music coding in the {Opus} codec,'' \emph{Journal of the Audio
  Engineering Society}, October 2013.

\bibitem{vos2013voice}
K.~Vos, K.~V. S{\o}rensen, S.~S. Jensen, and J.-M. Valin, ``Voice coding with
  {Opus},'' in \emph{Audio Engineering Society Convention}.\hskip 1em plus
  0.5em minus 0.4em\relax Audio Engineering Society, 2013.

\bibitem{makhoul1975lpc}
J.~Makhoul, ``Linear prediction: A tutorial review,'' \emph{Proceedings of the
  IEEE}, vol.~63, no.~4, pp. 561--580, 1975.

\bibitem{valin2009full}
J.-M. Valin, T.~B. Terriberry, and G.~Maxwell, ``A full-bandwidth audio codec
  with low complexity and very low delay,'' in \emph{European Signal Processing
  Conference (EUSIPCO)}.\hskip 1em plus 0.5em minus 0.4em\relax IEEE, 2009.

\bibitem{moore2012introduction}
B.~C.~J. Moore, \emph{An introduction to the psychology of hearing}.\hskip 1em
  plus 0.5em minus 0.4em\relax Brill, 2012.

\bibitem{rafaely2004analysis}
B.~Rafaely, ``Analysis and design of spherical microphone arrays,'' \emph{IEEE
  Transactions on speech and audio processing}, vol.~13, no.~1, pp. 135--143,
  2004.

\bibitem{li2005hemispherical}
Z.~Li and R.~Ruraiswami, ``Hemispherical microphone arrays for sound capture
  and beamforming,'' in \emph{Workshop on Applications of Signal Processing to
  Audio and Acoustics (WASPAA)}.\hskip 1em plus 0.5em minus 0.4em\relax IEEE,
  2005.

\bibitem{venkatesan2004iterative}
S.~Venkatesan, L.~Mailaender, and J.~Salz, ``An iterative algorithm for
  computing a spatial whitening filter,'' in \emph{Workshop on Signal
  Processing Advances in Wireless Communications (SPAWC)}.\hskip 1em plus 0.5em
  minus 0.4em\relax IEEE, 2004.

\bibitem{yang2003high}
D.~Yang, H.~Ai, C.~Kyriakakis, and C.-C. Kuo, ``High-fidelity multichannel
  audio coding with {Karhunen-Loeve} transform,'' \emph{IEEE Transactions on
  Speech and Audio Processing}, vol.~11, no.~4, pp. 365--380, 2003.

\bibitem{Compernolle1990}
D.~V. Compernolle, W.~Ma, F.~Xie, and M.~V. Diest, ``Speech recognition in
  noisy environments with the aid of microphone arrays,'' \emph{Speech
  Communication}, vol.~9, no.~5, pp. 433 -- 442, 1990.

\bibitem{heymann2016neural}
J.~Heymann, L.~Drude, and R.~Haeb-Umbach, ``Neural network based spectral mask
  estimation for acoustic beamforming,'' in \emph{International Conference on
  Acoustics, Speech and Signal Processing (ICASSP)}.\hskip 1em plus 0.5em minus
  0.4em\relax IEEE, 2016.

\bibitem{erdogan2016improved}
H.~Erdogan, J.~R. Hershey, S.~Watanabe, M.~I. Mandel, and J.~Le~Roux,
  ``Improved {MVDR} beamforming using single-channel mask prediction
  networks,'' in \emph{Interspeech}, 2016.

\bibitem{zhang2012deep}
X.-L. Zhang and J.~Wu, ``Deep belief networks based voice activity detection,''
  \emph{IEEE Transactions on Audio, Speech, and Language Processing}, vol.~21,
  no.~4, pp. 697--710, 2012.

\bibitem{Souden2013}
M.~Souden, S.~Araki, K.~Kinoshita, T.~Nakatani, and H.~Sawada, ``A multichannel
  {MMSE}-based framework for speech source separation and noise reduction,''
  \emph{IEEE Transactions on Audio, Speech, and Language Processing}, vol.~21,
  no.~9, pp. 1913--1928, Sept 2013.

\bibitem{Park2020}
\BIBentryALTinterwordspacing
D.~S. Park, Y.~Zhang, C.-C. Chiu, Y.~Chen, B.~Li, W.~Chan, Q.~V. Le, and Y.~Wu,
  ``{SpecAugment} on large scale datasets,'' \emph{International Conference on
  Acoustics, Speech and Signal Processing (ICASSP)}, May 2020. [Online].
  Available: \url{http://dx.doi.org/10.1109/ICASSP40776.2020.9053205}
\BIBentrySTDinterwordspacing

\end{thebibliography}
\end{document}